\documentclass [pra, twocolumn, aps, superscriptaddress, showpacs] {revtex4}
\usepackage{graphicx}
\usepackage{amsmath}
\usepackage{amssymb}

\DeclareMathOperator{\tr}{Tr}
\DeclareMathOperator{\sign}{sign}
\newcommand{\ket}[1]{|#1\rangle}

\begin{document}

\title{Fulde-Ferrell-Larkin-Ovchinnikov(FFLO) vs Bose-Fermi mixture 
in polarized \\1D Fermi gas on a Feshbach resonance: a 3-body study}

\author{Stefan K. Baur}
\email{skb37@cornell.edu}
\affiliation{Laboratory for Atomic and Solid State Physics, Cornell University, Ithaca, New York 14853, USA}
\author{John Shumway}
\affiliation{Department of Physics, Arizona State University, Tempe, Arizona 85287-1504, USA}
\affiliation{Laboratory for Atomic and Solid State Physics, Cornell University, Ithaca, New York 14853, USA}
\author{Erich J. Mueller}
\affiliation{Laboratory for Atomic and Solid State Physics, Cornell University, Ithaca, New York 14853, USA}

\date{\today}

\begin{abstract}
We study the three-fermion problem 
within
a 1D model of a Feshbach
resonance in order to gain insight into how the Fulde-Ferrell-Larkin-Ovchinnikov(FFLO)-like state
at small negative scattering lengths evolves into a Bose-Fermi
mixture at small positive scattering lengths. The FFLO state
possesses an oscillating superfluid correlation function, while
in a Bose-Fermi mixture correlations are monotonic. We find that this
behavior is already present at the three-body level. We present an
exact study of the three-body problem, and gain extra insights by considering world lines of a path-integral Monte Carlo calculation.
\end{abstract}


\maketitle
\section{Introduction}
Trapped ultracold clouds of fermions such as $^{6}\text{Li}$ 
provide unique insights into the superfluidity of neutral fermions and have opened up new directions for 
inquiry.  By considering the three-body problem, here we theoretically address the properties of a 
one-dimensional (1D)
superfluid gas of spin-imbalanced fermions (where $n_\uparrow>n_\downarrow$) when the interactions are tuned via a Feshbach resonance. We find a change in symmetry of the ground-state wave function as a function of system parameters, and connect this symmetry change with properties of the many-body state. Our conclusions come from (i) the scattering lengths calculated from an exact solution of the 3-body problem and (ii) the off-diagonal elements of the pair density matrix calculated with path-integral Monte Carlo. In the latter formulation the symmetry change in the wave function emerges from a competition between two classes of topologically distinct imaginary-time world lines. Our conclusions are relevant to  experiments on $^6$Li atoms trapped in an array of very elongated traps, formed from a two-dimensional optical lattice \cite{moritz:210401}.  When such a
lattice is sufficiently strong, one has an array of independent 1D systems,
and experiments probe ensemble-averaged quantities including the momentum distribution of pairs. 

Similar experiments in three dimensions (3D)
 have demonstrated  a
crossover between BCS superfluidity of loosely bound pairs to
a Bose-Einstein condensation (BEC) of molecules, finding particularly 
rich physics (mostly involving phase separation) when the gas is
 spin polarized~\cite{spinpolarizedexp}.  
 One dimension brings a new set of phenomena, driven by 
  quantum fluctuations and the topology of the Fermi surface.

Of particular interest, Fermi surface nesting in 1D  stabilizes~\cite{1dpolarizedtheory} a version of the Fulde-Ferrell-Larkin-Ovchinnikov (FFLO) phase in the spin-imbalanced gas \cite{fflo}.  FFLO phases, which occupy an extremely small region of the 3D phase diagram~\cite{3dfflo}, are characterized by a coexistence of magnetic and superfluid order, typically coupled together with a spin-density wave.
An intuitive example is given by a quasi-1D spin-imbalanced BCS superfluid,
where one finds an array of $\pi$-domain walls in the superfluid order parameter, with the excess unpaired atoms
 residing near the nodes \cite{1dfflomft}. At higher polarizations the domain walls merge, and the order parameter becomes sinusoidal.
We are interested in the truly 1D limit, where there is no long range order: Instead, one can introduce an operator $b(x)$ which annihilates a pair at position $x$, finding the analogy of FFLO state is that
$\left<b^{\dagger}(x)b(0)\right>\sim \cos(2 \pi n_F x)/|x|^{\delta}$  where $n_F=n_{\uparrow}-n_{\downarrow}$ is the density of excess fermions and the exponent $\delta$ depends on interactions \cite{1dfflobeyondmft}.

When the interactions are weak,
a sufficiently dilute and cold gas of $^6$Li atoms in an elongated trap (with transverse dimension $d=\sqrt{\hbar/m\omega_{\perp}}$) can be modeled as a 1D Fermi gas interacting through a short-range 1D potential \cite{olshanii1998}.  
This mapping requires that the 3D scattering length is negative with $|a|/d\ll1$,
and both the thermal energy $k_B T$ and the chemical potential $\mu$ are small compared to the transverse confinement energy $\hbar\omega_{\perp}$.  Like Refs. \cite{1dbecbcscrossover,mora3body}, we will consider stronger interactions.
The breakdown of the mapping onto a 1D Fermi gas is illustrated by the situation where the 3D scattering length is small and positive, hence producing a deeply bound molecular state.  
The correct description of the unpolarized system in this limit is a weakly interacting gas of these bosons: a model which is not equivalent to a 1D gas of fermions with point interactions.

If one spin imbalances the system in this BEC limit, one does not produce a FFLO state, but rather the excess fermions only mildly perturb the bosonic pairs, and the correlation function $\left<b^{\dagger}(x)b(0)\right>\sim 1/|x|^{\delta'}$ is monotonic \cite{imambekov:021602}.
Here we study the three-body problem to address the key question of how a spin-imbalanced gas evolves between this fluctuating ``BEC" limit and the fluctuating ``BCS" limit already described.
How does the correlation function go from monotonic to oscillatory? We find that in the three-body problem the transition occurs due to a level crossing.

To this end, we consider
the minimal 1D model of a Feshbach resonance~\cite{recati:033630,sachdev2006}, which can capture the relevant physics,
\begin{eqnarray}
\label{bfhamiltonian}
H&=&\sum_{k,\sigma} \frac{\hbar^2 k^2}{2 m} c_{k,\sigma}^{\dagger}c_{k,\sigma}+\sum_{k} \left( \frac{\hbar^2 k^2}{4 m}+\nu \right) b_k^{\dagger} b_k\\
&+&\frac{g}{\sqrt{L}} \sum_{q,Q} b_Q^{\dagger} c_{Q/2+q,\downarrow} c_{Q/2-q,\uparrow}+h.c.\nonumber,
\end{eqnarray}
where $L$ is the length of the system and $c^{\dagger}_{k,\sigma}$, $c_{k,\sigma}$($b^{\dagger}_{k}$, $b_{k}$) are fermionic(bosonic) creation/annihilation operators.
The parameter $g$ describes the coupling strength between the bosonic and fermionic channel and $\nu$ is the detuning with $\nu\rightarrow \infty$ ($\nu\rightarrow -\infty$) being the BCS (BEC) limit.
We will use units in which $\hbar^2/m=1$. \footnote{Following submission of this paper a numerical study of the manybody
problem of a similar model was carried out by the authors of
[F. Heidrich-Meisner, A. E. Feiguin, U. Schollw\"ock, and W. Zwerger,
Phys. Rev. A \textbf{81}, 023629 (2010)] finding results consistent with those
reported here.}

\begin{figure}
	\centering
		\includegraphics[width=\linewidth]{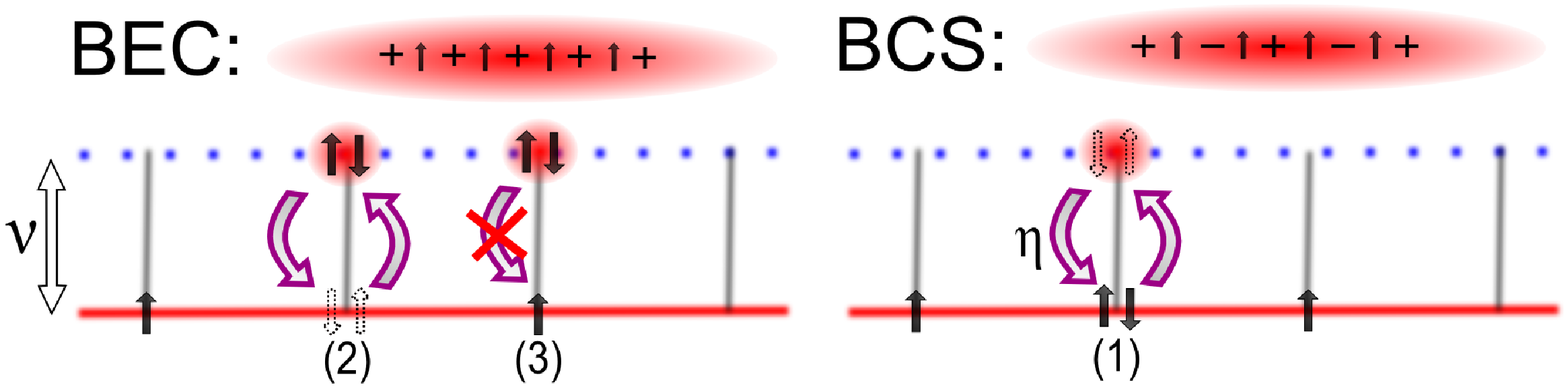}
		\includegraphics[width=\linewidth]{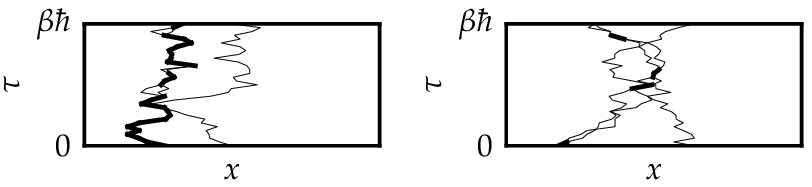}
	\caption{(Color online) Cartoon depictions of the physics of Eq.~(\ref{bfhamiltonian}) in the BEC (left) and BCS (right) limits.  
	(Top) Symmetry of Bose wave function: in the BCS limit the wave function changes sign whenever a pair passes a (spin-up) fermion. (Middle) Depiction of lattice model which is used for developing intuition about Eq. (1). (Bottom) Typical world lines illustrating interaction of a boson (heavy line) and fermion (thin line) with space along the horizontal axis and imaginary-time along the vertical axis.
}
	\label{fig:laddercartoon}
\end{figure}

\section{Qualitative Structure}
Figure ~\ref{fig:laddercartoon} shows a cartoon depiction of the lattice version of this model.  
One can represent the model in terms of two 1D channels, represented 
as the legs of a ladder.  Fermions move on the lower leg, while 
bosons move on the upper.  As shown at (1) and (2), pairs of fermions can hop from the lower leg to the upper, becoming a boson and vice versa.   

In the BCS limit, 
$\nu\gg g^{4/3}$, the atoms mainly sit on the lower leg, making virtual transitions to the bosonic leg.  These virtual transitions lead to a weak local attraction between fermions, $U=-g^2/\nu$.  The 
figure on the bottom right illustrates typical world lines for three fermions.

In the BEC limit, 
 $-\nu\gg g^{4/3}$, the atoms mainly sit on the upper leg.  They make virtual transitions to the lower leg.
 As illustrated at (3), a boson cannot make a virtual transition if an excess fermion sits at that location.  This 
 leads to a repulsive interaction between the bosons and fermions of strength $g^2/\nu$.  Unlike the BCS limit, the world-lines of the fermions and bosons cross.  

 \begin{figure}
	\centering
		\includegraphics[width=\columnwidth]{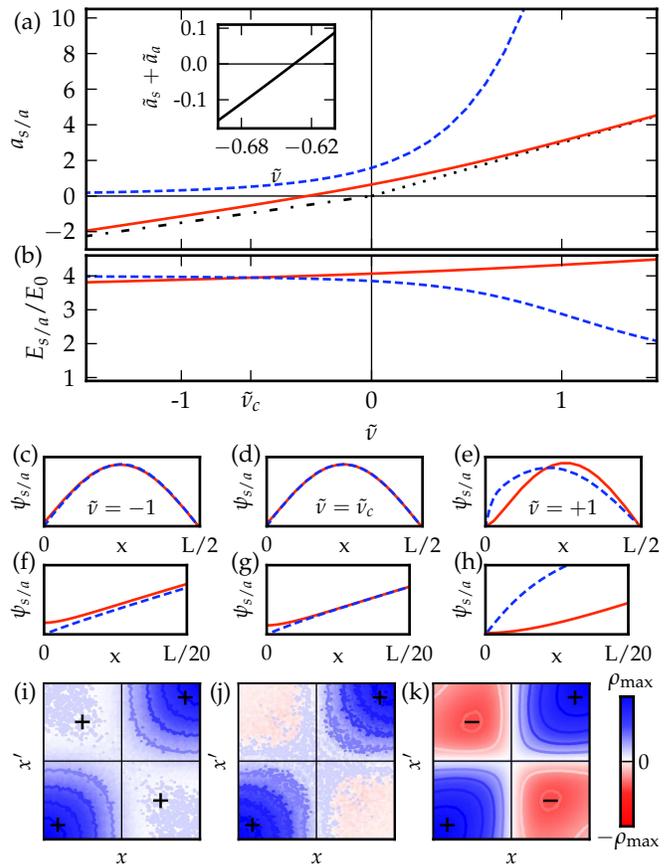}
		\caption{(Color online) The dimensionless 1D scattering lengths $\tilde{a}_{s/a}=a_{s/a} g^{2/3}$ for the symmetric (solid red line)/antisymmetric (dashed blue line) channel plotted vs the dimensionless detuning $\tilde{\nu}=\nu/g^{4/3}$. The dotted (dashed dotted) line is the asymptotic result for $a_s$, $a_s=3 \nu/g^2$ ($a_s=(3/2) \nu/g^2$) in the BCS limit (BEC limit); (cf \cite{mora3body}). (Inset) Sum $\tilde{a}_s+\tilde{a}_a$ (solid line) crosses zero at $\tilde{\nu}\approx-0.635$, marking the change in symmetry of the ground state.
		(c)-(e) Lowest-energy symmetric (solid line)/antisymmetric (dashed line) wave function $f_{s/a}(x)=L^{-1/2} \sum_Q e^{i Q x} f_{s/a,Q}$, in a hard-wall box of size $L \approx 160/g^{2/3}$, where $x$ represents the relative separation of the boson and fermion.
		 Left to right: $\tilde{\nu}=-1,-0.635,1$.
		 (f)-(h) Wave function near the origin.  Finite range of the effective interaction is apparent from the nonsinusoidal shape of $f$ for small $x$.  (i)-(k) Reduced density-matrix $\rho(x,x^\prime)$ defined in the text before Eq.~(\ref{eq:rho}) for $\beta=100/g^{4/3}$ caculated with QMC.  Blue/red represents positive/negative weight.  Quadrants with predominant positive/negative weight are labeled with ``+''/``--''.   
		  }
	\label{fig:wavefunctions}
\end{figure}
\section{Wave functions}
 To gain insight into how this symmetry change occurs, we study the eigenstates of 
 Eq.~(\ref{bfhamiltonian}) for the case of 
 three particles.  Mora \textit{et al.}~\cite{mora3body} carried out a similar study for a more sophisticated model
 of fermions confined to a harmonic waveguide.  The simpler nature of our model, which only includes the most relevant degrees of freedom, makes the physics more transparent.  

 We study what the symmetry of the ground state is as a function of the dimensionless parameter $\tilde{\nu}=\nu/g^{4/3}$.  
 Given that the three-body wave function can be written
\begin{equation}
\begin{split}
\ket{\Psi}=\Big(& \sum_{K} f_K b_K^{\dagger} c_{-K,\uparrow}^{\dagger}\\
+&\sum_{k,K} g_{K,k} c_{K,\downarrow}^{\dagger} c_{k-K/2,\uparrow}^{\dagger} c_{-k-K/2,\uparrow}^{\dagger}\Big)\ket{0},
\end{split}
\end{equation}
we ask what the symmetry of $f_K$ is under switching the relative position of the boson and the fermion (i.~e. $K\to-K$).  We find that the ground state $f$ switches from odd (consistent with FFLO) to even (consistent with a Bose-Fermi mixture) as $\nu$ is increased from large negative values.   

To arrive at this result, we integrate out the three-fermion part of the wave function~\cite{3bodyfeshbach}, deriving an integral equation for the two-particle wave function $f_K$, 
\begin{equation}
\mathcal{L}(Q,E) f_Q=-\frac{g^2}{L}\sum_{K'} \frac{f_{K'}}{K'^{2}+Q K'+Q^2-E},
\end{equation}
where 
\begin{equation}
\mathcal{L}(Q,E)=3 Q^2/4+\nu-E-g^2/(2 \sqrt{3 Q^2/4-E})
\end{equation}
For details, see Appendix \ref{appendixa}.

The low-energy symmetric and antisymmetric scattering states have the form $\psi_s(x) \propto \sin[k (|x|-a_s)]$ and $\psi_a(x) \propto \sin[k (x+\sign(x)a_a)]$ for large $|x|$.  By imposing hard-wall boundary conditions, $f(x=\pm L/2)=0$, one sees that the ground state will be symmetric when $a_s>-a_a$, and antisymmetric otherwise.  Figure ~\ref{fig:wavefunctions}(a) shows  these scattering lengths as a function of $\nu$, revealing that 
the symmetry of the wave function changes at
$\tilde{\nu}\approx-0.635$
, where the two solutions are degenerate.  Figures ~\ref{fig:wavefunctions}(b)-(h) shows the structure of the lowest energy symmetric and antisymmetric wave functions with these boundary conditions.    Note that on the BCS side of resonance, where $-a_a>a_s$, the Bose-Fermi interaction cannot be described by a local potential, rather it is a more general kernel~\cite{mora3body}.  The off-diagonal nature of the interaction
 allows the system to violate the standard theorem that the ground-state wave function of a nondegenerate system has no nodes.
The level crossing between the states of differing symmetry suggests one of several scenarios for the many-body system, with the most likely candidates being a first-order phase transition or a crossover.    Similar behavior was seen by Kestner and Duan \cite{kestner:033611} in their investigation of the three-body problem in a 3D harmonic trap.

\section{Quantum Monte Carlo (QMC)}
We developed a QMC algorithm to calculate thermodynamic quantities in this model and to give alternative ways of thinking about the underlying physics.
We calculate the thermal density matrix
\begin{equation}
\rho(x,x')=Z^{-1}\tr \left[ e^{-\beta H} b^{\dagger}(x) c_{\uparrow}^{\dagger}(0) c_{\uparrow}(0) b(x') \right],
\end{equation}
where $b(x)=L^{-\frac{1}{2}}\sum_k e^{i k x} b_k$, 
$c_{\sigma}(x)=L^{-\frac{1}{2}}\sum_k e^{i k x} c_{k,\sigma}$, 
$\beta=1/k_B T$, and $Z$ 
is the partition function. Figures ~\ref{fig:wavefunctions}(i)-(k) 
shows a density plot of this correlation function.  
The FFLO phase is distinguished from the Bose-Fermi mixture 
by the sign of $\rho$ in the top left and bottom right quadrants. 
The boundary between these behaviors occurs roughly where 
$-a_a=a_s$.

Considering 
first the fermionic sector, with two spin-up and one spin-down fermions,
 we discretize imaginary time into ${\cal N}$ slices, writing 
\begin{equation}
\label{eq:rho}
\begin{split}
&\rho(
{x_{\mathcal{\bar{N}}}
^{1\uparrow}
},
{x_{\mathcal{\bar{N}}}^{2\uparrow}},
x_\mathcal{N}^\downarrow;\;
x_0^{1\uparrow},x_0^{2\uparrow},x_0^\downarrow;\;\beta)=
\frac{1}{2Z}\\
&\left[\int_I \prod_j  dx^{1\uparrow}_j dx^{2\uparrow}_j dx^{\downarrow}_j
e^{-S}-\int_X \prod_j  dx^{1\uparrow}_j dx^{2\uparrow}_j dx^{\downarrow}_j
e^{-S}\right]
\end{split}
\end{equation}
as integrals over the positions of the up-spins $x^{i\uparrow}$ and the down-spin $x^\downarrow$ 
at imaginary times $\tau_j=j\beta/\mathcal{N}$, with discretized action $S$. 
For appropriately chosen $S$, this expression converges to the exact thermal expectation value as ${\cal N}\to\infty$.  
Two separate boundary conditions account for the fermionic statistics:
$\int_I$ has $x_\mathcal{N}^{1\uparrow}={x_\mathcal{\bar{N}}^{1\uparrow}}$
and $x_\mathcal{N}^{2\uparrow}={x_\mathcal{\bar{N}}^{2\uparrow}}$ while
$\int_X$ has $x_\mathcal{N}^{1\uparrow}={x_\mathcal{\bar{N}}^{2\uparrow}}$
and $x_\mathcal{N}^{2\uparrow}={x_\mathcal{\bar{N}}^{1\uparrow}}$
The integrals are performed by a Monte Carlo algorithm, 
treating $e^{-S}$ as a probability measure.  Details of our choice
of discretized action and the resulting Monte Carlo rules are given
in Appendix~\ref{sec:qmcrules}.

\begin{figure}
\centering
\includegraphics[width=\linewidth]{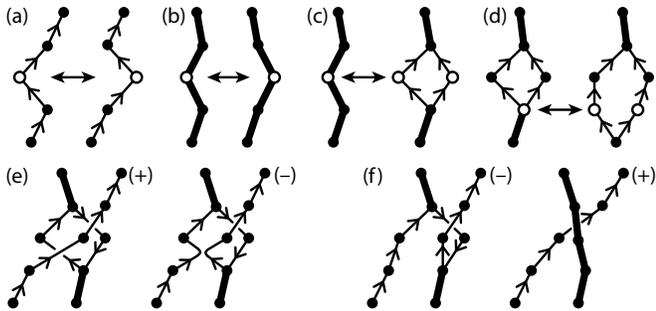}
\caption{Illustrative moves in our QMC algorithm. Fermions are designated by thin lines with arrows representing the spin, bosons by thick lines, and moving beads are white: (a) Moving a fermion, (b) moving a boson, (c) opening/closing, and (d) zipping/unzipping. (e) Crossing of same-spin fermions is always canceled by an equal weight path of opposite sign. (f) Bosons enable paths with both negative and positive weight that do not cancel.}
\label{moves}
 \end{figure}

While path-integral QMC techniques are well established \cite{RevModPhys.67.279}, the present situation is novel because two fermions can bind and form a boson.  We implement this feature by introducing extra variables that record the slices at which two fermions are bound, and requiring that when two fermions are bound (say $x_j^{1\uparrow}$ and $x_j^{\downarrow}$)
 then their positions must be equal.
The moves in our Markov process are as follows: moving a particle in one time slice, binding two unbound fermions of opposite spin into a boson,
and unbinding two fermions. In all cases  the probabilities of the move in slice $j$ only depends
on the positions at time slices $j-1$ and $j+1$.  Sampling new positions 
from a Gaussian centered about weighted average of the particle's position in the previous
and last slice optimizes the acceptance rate. 
As described in Appendix~\ref{sec:qmcrules},
we find the rules summarized in Table~\ref{table:mc}
and illustrated in Figs.~\ref{moves}(a)-(d),
which let \eqref{eq:rho} converge to the exact density matrix as $\mathcal{N}\to\infty$. Specifying these Markov rules is equivalent to specifying $S$.

\begin{table*}
\begin{ruledtabular}
\caption{Gaussian sampling widths and Metropolis acceptance rule,
 $\mathcal{A}=\min(1,e^{-\Delta S} T_R/T_F)$, for moves in Figs.~\ref{moves} (a)-(d).
 Moves for bead $x_j'\rightarrow x_j$ are sampled from a 
Gaussian of width $\sigma_F$
 centered about $\bar x_j$;
while the reverse moves $x_j\rightarrow x_j'$
 sample a Gaussian of width $\sigma_R$.
 \label{table:mc}}
\begin{tabular}{lcccr}
Move & Sampling  midpoint $\bar{x}_j$
& Sampling width $\sigma_F$ 
& Sampling width $\sigma_R$ & $e^{-\Delta S} T_R/T_F$ \\
\hline
(a) Fermion 
 & $\frac{x^\uparrow_{j+1}-x^\uparrow_{j-1}}{2}$
 & $\sqrt{\Delta\tau/2}$
 & $\sqrt{\Delta\tau/2}$
  & 1\\
(b) Boson 
 &$\frac{x_{j+1}-x_{j-1}}{2}$
 & $\sqrt{\Delta\tau/4}$
 & $\sqrt{\Delta\tau/4}$
 & 1\\
(c) Close$\rightarrow$open 
&$\frac{x_{j+1}-x_{j-1}}{2}$
& $\sqrt{\Delta\tau/2}$
& $\sqrt{\Delta\tau/4}$
&$\exp(\nu\,\Delta\tau)\Big/g^2\,\Delta\tau^2\sqrt{8\pi\,\Delta\tau}$
\\
(d) Zip$\rightarrow$unzip 
 &$\frac{x^\uparrow_{j+1}+x^\uparrow_{j+1}-2x_{j-1}}{4}$
 & $\sqrt{\Delta\tau/2}$
 &$\sqrt{\Delta\tau/4}$
 &$ \exp\left(\nu\Delta\tau + \frac{\left|x^\uparrow_{j+1}
 -x^\downarrow_{j+1}\right|^2}{8\Delta\tau}\right)\big/\sqrt{2}
$
  \\
\end{tabular}
\end{ruledtabular}
\end{table*}

 Since the density matrix involves adding up terms with different signs, at low temperatures or
 large particle numbers the efficiency can suffer; this is the  ``fermion sign problem."  For three particles the variance remains small enough that we can produce reasonably accurate results with the algorithm already described.  To make further improvements, 
 we use of the fact that paths cancel when world lines for identical fermions cross in 1D, a well-known technique
for eliminating the sign problem in 1D\footnote{Note that our model is not purely 1D as the boson channel provides a mechanism for fermions to move past one-another.}.
 For example, Fig.~\ref{moves}(e) illustrates two paths for which $e^{-S}$ has the same value, but which contribute to $\rho$ with opposite signs.  We therefore throw away both sets of paths.  In a purely 1D system of fermions one could thereby eliminate all paths with one sign or the other, depending on the relative ordering of the particles at the beginning and end.
  Here the cancellation is incomplete.  Figure ~\ref{moves}(f) illustrates  paths of opposite sign which have no
  term of the opposite sign to cancel.
 When the exchanges are dominated by paths with positive weights (such as the RHS of Fig. 3(f)) one has a Bose-Fermi mixture; otherwise one has an FFLO-like state.
  
\section{Realization/Detection}
We studied the simplest model for the BEC-BCS crossover of spin-polarized fermions in harmonic waveguides, a many-body system realizable by spin imbalancing the array of 1D tubes created in~\cite{moritz:210401}.  In such an experiment one could distinguish FFLO from a Bose-Fermi mixture by either using an interferometric probe~\cite{gritsev-2008} or measuring the pair momentum distribution, e.g. by sweeping to the BEC side followed by time-of-flight expansion. The signature of the FFLO phase is a peak at finite momentum $q=\pi n_F$ set by the density of excess fermions $n_F=n_{\uparrow}-n_{\downarrow}$~\cite{casula2008}. This peak should be absent in a Bose-Fermi mixture with monotonically decaying superfluid correlations. Another probe, based on correlations in the atomic shot noise after time-of-flight expansion, has been suggested in~\cite{luescher2008}. Additionally, there has recently been effort in studying the BEC-BCS crossover in few-body clusters~\cite{gemelke2008}. 
By creating ensembles of elongated clusters one can directly realize and study the three-body system considered here: tuning interactions using a photoassociation or a Feshbach resonance~\cite{recati:033630,harmonicwaveguide}.


\acknowledgments
We would like to thank K. Hazzard, D. Huse, W. Zwerger and R. Hulet for useful discussions. This work was supported under ARO Award W911NF-07-1-0464 with funds from the DARPA OLE program and used computer resources at the Cornell Nanoscale Facility, a member of the NSF supported National Nanotechnology Infrastructure Network.

\appendix
\section{Solution of the 3-body problem}
\label{appendixa}
Applying the Hamiltonian Eq. \ref{bfhamiltonian} to $\ket{\Psi}$ gives a pair of coupled Schr\"odinger equations
\begin{eqnarray*}
\left(k^2+\frac{3}{4} K^2-E \right) g_{K,k}-\frac{g \left(f_{k+K/2}-f_{-k+K/2}\right)}{2\sqrt{L}}=0 \\
\left(\frac{3}{4}K^2+\nu-E\right) f_K-2 \frac{g}{\sqrt{L}} \sum_k g_{K/2+k,3 K/4-k/2}=0.
\end{eqnarray*}
In the first equation we antisymmetrized the second term in $k$ to ensure manifest antisymmetry of $g_{K,k}$. Eliminating $g_{K,k}$ from the last equation gives an integral equation for $f_K$
\begin{eqnarray}
\left(\frac{3 }{4}K^2+\nu-E\right) f_K=\frac{g^2}{L} \sum_k \frac{f_K-f_{k-K/2}}{k^2+3 K^2/4-E}
\end{eqnarray}
After performing the integral $\int_{-\infty}^{\infty} dk/\left[2 \pi(k^2+3 K^2/4-E)\right]=1/(2 \sqrt{3 K^2/4-E})$ this simplifies to (we relabeled $K$ to $Q$)
\begin{eqnarray}
\label{inteq}
\mathcal{L}(Q,E) f_Q=-\frac{g^2}{L} \sum_{K'} \frac{f_{K'}}{K'^{2}+Q K'+Q^2-E}
\end{eqnarray}
with $\mathcal{L}(Q,E)=3 Q^2/4+\nu-E-g^2/(2 \sqrt{3 Q^2/4-E})$. The solution $E_B$ of the equation $\mathcal{L}(0,E_B)=0$ is the two-body bound state energy ~\cite{recati:033630},\footnote{Solving $L(0,E_B)=0$ gives $E_B=\nu/(3 \xi)-\xi$ with $\xi=\left(27 g ^4+8 \nu ^3+3 g^2 \sqrt{81 g ^4+48 \nu ^3}\right)^{1/3}$. The asymptotics are $E_B\sim \nu$ for $\nu\rightarrow-\infty$(BEC limit) and $E_B\sim -g^4/4 \nu^2$ for $\nu\rightarrow+\infty$(BCS limit) ~\cite{recati:033630}.} and $\mathcal{L}(Q,E_Q)=0$ is solved by $E_Q=3 Q^2/4+E_B$.
Equation ~(\ref{inteq}) can be converted into a Lippmann-Schwinger equation for the scattering amplitude $F(Q,K)$ using the ansatz~\cite{mora3body},
\begin{equation}
f_Q=2\pi \delta(K-Q)+i F(Q,K) \sum_{\pm} 1/(K\pm Q+i\epsilon),
\end{equation}
which gives
\begin{equation}
\begin{split}
\label{ls}
2 i F(Q,K) K=&-V(Q,K,K)\\
&-i\frac{1}{L}\sum_{K',\pm} \frac{F(K',K)}{K\pm K'+i \epsilon} V(Q,K',K)
\end{split}
\end{equation}
where we introduced the effective boson fermion potential
\begin{equation}
V(Q,K',K)=g^2 \frac{K^2-Q^2}{\mathcal{L}(Q,E_K)(K'^{2}+Q K'+Q^2-E_K)}
\label{effint}
\end{equation}
At low energies(small momentum $K$\footnote{What we mean here is that the typical size of a pair $r=1/\sqrt{-E_B}$ is much smaller than the interparticle spacing $r_i=1/n_p$($n_p$ is the density of pairs). In cold atom system not too far on the BCS side of the resonance, this condition can typically be achieved.}) one has $F(Q,K)\approx-1+i K a_s+i Q a_a$ ~\cite{mora3body} where $a_s$($a_a$) is the scattering length for the symmetric(antisymmetric) channel. To extract the low-energy scattering properties, we numerically solve the integral equation Eq.~(\ref{ls}) at fixed small $K$ and compute the scattering length from the limits 
\begin{align}
a_s&=\lim_{K \rightarrow 0} \text{Im}[F(K,K)+F(-K,K)]/2 K,\\
a_a&=\lim_{K \rightarrow 0} \text{Im}[F(K,K)-F(-K,K)]/2 K.
\end{align}


\section{Derivation of the path integral action and Monte Carlo rules}
\label{sec:qmcrules}

The partition function $Z$ corresponding to the Hamiltonian [Eq.~(\ref{bfhamiltonian})] can be expressed as a path integral.
The path-integral formulation is useful both as a computational tool, but also provides insights from a different point of view. We formulate the
path integral in real space (position basis) and imaginary time, $0\le \tau \le \beta$.
We discretize imaginary time into discrete steps $\Delta\tau = \beta/
\mathcal{N}$,
where $\mathcal{N}$ is the Trotter number.
The path integral is equivalent to Eq.~(\ref{bfhamiltonian}) in the limit 
$\mathcal{N}\rightarrow \infty$, which is taken by extrapolating our numerical results
to $\Delta\tau\rightarrow 0$, following the high-accuracy method
of Schmidt and Lee~\cite{schmidt1995} .

\begin{figure}[b]
\centering
\includegraphics[width=\linewidth]{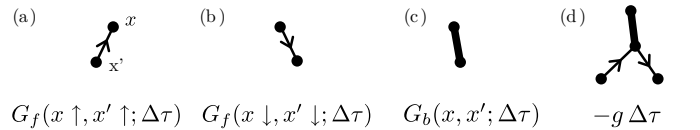}
\caption{Propagators and interaction vertex for disretized path-integral model of 
Eq.~(\ref{bfhamiltonian}). (a) Spin-up free fermion, Eq.~(\ref{eq:fprop}),
(b) spin-down free fermion, Eq.~(\ref{eq:fprop}),
(c) free boson, Eq.~(\ref{eq:bprop}),
and (d) interaction vertex with extra weight $-g\,\Delta\tau$. }
\label{fig:prop}
\end{figure}

 To construct the
path integral, we start from the imaginary time propagators for 
free fermions,
\begin{equation}
\begin{split}\label{eq:fprop}
G_f(x\sigma, x'\sigma';\Delta\tau)
&= \langle 0 | c_\sigma(x)
e^{-\Delta\tau\, H} c^\dagger_\sigma(x')|0\rangle\\
&= \delta_{\sigma\sigma'}\exp\left( -\frac{|x-x'|^2}{2 \Delta\tau}\right)
\Big/ \sqrt{2\pi\,\Delta\tau},
\end{split}
\end{equation}
and for free bosons,
\begin{equation}
\begin{split}\label{eq:bprop}
G_b(x, x';\Delta\tau)
&= \langle 0 | b(x)
e^{-\Delta\tau\, H} b^\dagger(x')|0\rangle\\
&= \exp\left( -\frac{|x-x'|^2}{\Delta\tau}-\nu\,\Delta\tau\right)
\Big/ \sqrt{\pi\,\Delta\tau}.
\end{split}
\end{equation}
These propagators are illustrated in Fig.~\ref{fig:prop}.
To represent the interaction, we weigh a vertex [Fig.~\ref{fig:prop}(d)] with
$-g\,\Delta\tau$.
The action $S$ for a path configuration, used in Eq.(~\ref{eq:rho}), is then
given by the negative of the log of the product of the propagators and 
interaction vertices that make up the path.

To sample the path, we use the Metropolis algorithm, in which the
acceptance of a move is given by
\begin{equation}\label{eq:metropolis}
\mathcal{A} = \min(1,e^{-\Delta S}T_R/T_F),
\end{equation}
where $T_F$ and $T_R$ are the forward and
reverse probabilities of attempting
a particular move. For example, consider the 
move illustrated in Fig.~\ref{moves}(c), where
a bead on a bosonic path is split to form a fermionic pair
(a bubble). For the forward move, we sample the two fermion
positions $x_j^\uparrow$ and
$x_j^\downarrow$
from a Gaussian of width $\sigma_F=\sqrt{\Delta\tau/2}$ centered
about $\bar{x}_j=(x_{j+1}+x_{j-1})/2$
where $x_{j-1}$ and $x_{j+1}$ are the stationary boson positions
immediately before and after the sampled slice.
For the reverse move, we sample the recombined boson
position $x_j$ from a Gaussian of width $\sigma_R\sqrt{\Delta\tau/4}$
centered about $\bar{x}_j$.
We find
\begin{widetext}
\begin{equation}
\begin{split}
e^{-\Delta S}\frac{T_R}{T_F}&=
\frac{G_f(x_{j+1},x_j^\uparrow)G_f(x_j^\uparrow,x_{j-1})
      G_f(x_{j+1},x_j^\uparrow)G_f(x_j^\uparrow,x_{j-1})(g\,\Delta\tau)^2}
     {G_b(x_{j+1},x_j)G_b(x_j,x_{j-1})}
\frac{e^{-\frac{|x_j-\bar{x}_j|^2}{2\sigma_R}}(2\pi\sigma_R^2)^{-\frac{1}{2}}}
     {e^{-\frac{|x_j^\uparrow-\bar{x}_j|^2}{2\sigma_F}}
      e^{-\frac{|x_j^\downarrow-\bar{x}_j|^2}{2\sigma_F}}
      (2\pi\sigma_F^2)^{-1}} \\
&=\frac{\exp(\nu\,\Delta\tau)}{g^2\,\Delta\tau^2\sqrt{8\pi\,\Delta\tau}}.
\end{split}
\end{equation}
\end{widetext}
This rule and the rules for the other moves illustrated 
in Figs.~\ref{moves}(a)-(d) are summarized in
Table~\ref{table:mc}.


\begin{thebibliography}{10}

\bibitem{moritz:210401}
H. Moritz {\it et~al.}, Phys. Rev. Lett. {\bf 94},  210401  (2005).

\bibitem{spinpolarizedexp}
G.~B. Partridge {\it et~al.}, Science {\bf 311},  503  (2006); M.~W. Zwierlein {\it et~al.}, Science {\bf 311},  492  (2006).

\bibitem{1dpolarizedtheory}
G. Orso, Phys. Rev. Lett. {\bf 98},  070402  (2007); H. Hu, X.-J. Liu, and P.~D. Drummond, Phys. Rev. Lett. {\bf 98},  070403
  (2007); M.~M. Parish {\it et~al.}, Phys. Rev. Lett. {\bf 99},  250403  (2007).

\bibitem{fflo}
P. Fulde and R.~A. Ferrell, Phys. Rev. {\bf 135},  A550  (1964); A.~I. Larkin and Y.~N. Ovchinnikov, Sov. Phys. JETP {\bf 20},  762  (1965).

\bibitem{3dfflo}
D.~E. Sheehy and L. Radzihovsky, Phys. Rev. Lett. {\bf 96},  060401  (2006); M.~M. Parish {\it et~al.}, Nat. Phys. {\bf 3},  124  (2007).

\bibitem{1dfflomft}
A.~I. Buzdin and V.~V. Tugushev, Sov. Phys. JETP {\bf 58},  428  (1983); K. Machida and H. Nakanishi, Phys. Rev. B {\bf 30},  122  (1984).

\bibitem{1dfflobeyondmft}
K. Yang, Phys. Rev. B {\bf 63},  140511  (2001);
M. Rizzi {\it et~al.}, Phys. Rev. B {\bf 77},  245105  (2008);
A.~E. Feiguin and F. Heidrich-Meisner, Phys. Rev. B {\bf 76},  220508  (2007);
M. Tezuka and M. Ueda, Phys. Rev. Lett. {\bf 100},  110403  (2008);
E. Zhao and W.~V. Liu, Phys. Rev. A {\bf 78},  063605  (2008).

\bibitem{olshanii1998}
M. Olshanii, Phys. Rev. Lett. {\bf 81},  938  (1998).

\bibitem{1dbecbcscrossover}
J.~N. Fuchs, A. Recati, and W. Zwerger, Phys. Rev. Lett. {\bf 93},  090408
  (2004);
I.~V. Tokatly, Phys. Rev. Lett. {\bf 93},  090405  (2004);
C. Mora {\it et~al.}, Phys. Rev. Lett. {\bf 95},
  080403  (2005);
D. Blume and D. Rakshit, Phys. Rev. A 80, 013601  (2009).

\bibitem{mora3body}
C. Mora {\it et~al.}, Phys. Rev. Lett. {\bf 93},
  170403  (2004);
C. Mora, R. Egger, and A.~O. Gogolin, Phys. Rev. A {\bf 71},  052705  (2005).

\bibitem{imambekov:021602}
A. Imambekov and E. Demler, Phys. Rev. A {\bf 73},  021602  (2006).

\bibitem{recati:033630}
A. Recati, J.~N. Fuchs, and W. Zwerger, Phys. Rev. A {\bf 71},  033630  (2005).

\bibitem{sachdev2006}
D. E. Sheehy and L. Radzihovsky, Phys. Rev. Lett. {\bf 95}, 130401 (2005);
R. Citro and E. Orignac, Phys. Rev. Lett. {\bf 95}, 130402 (2005);
S. Sachdev and K. Yang, Phys. Rev. B {\bf 73}, 174504 (2006).

\bibitem{3bodyfeshbach}
A.~O. Gogolin, C. Mora, and R. Egger, Phys. Rev. Lett. {\bf 100},  140404
  (2008);
M. Jona-Lasinio, L. Pricoupenko, and Y. Castin, Phys. Rev. A {\bf 77},  043611
  (2008).


\bibitem{kestner:033611}
J.~P. Kestner and L.-M. Duan, Phys. Rev. A {\bf 76},  033611  (2007).

\bibitem{RevModPhys.67.279}
D.~M. Ceperley, Rev. Mod. Phys. {\bf 67},  279  (1995).

\bibitem{gritsev-2008}
V. Gritsev, E. Demler, and A. Polkovnikov, Phys. Rev. A {\bf 78},  063624
  (2008).

\bibitem{casula2008}
M. Casula, D.~M. Ceperley, and E.~J. Mueller, Phys. Rev. A {\bf 78}, 033607 (2008).

\bibitem{luescher2008}
A. L\"uscher, R.~M. Noack and A.~M. L\"auchli, Phys. Rev. A {\bf 78}, 013637 (2008).

\bibitem{gemelke2008}
N. Gemelke {\it et~al.}, in \textit{Proceedings of the XXI International Conference on Atomic Physics, Storrs,  2008}, edited by R. C\^ot\'e, P. L. Gould, M. Rozman and W.W. Smith(World Scientific, 2009), p. 240

\bibitem{harmonicwaveguide}
For a harmonic waveguide on a Feshbach resonance our argument that $a_s=-a_a$ for the symmetry change of the three-body wave function still holds. We estimate the transition at $d/a=1.7$\cite{mora3body}.

\bibitem{schmidt1995}
K.~E. Schmidt and M.~A. Lee, Phys. Rev. E {\bf 51}, 5495 (1995).

\end{thebibliography}
\end{document}